\documentclass{aa} 

\usepackage{graphicx}
\usepackage{txfonts}

\usepackage{booktabs}

\usepackage{natbib}
\usepackage{comment}
\bibpunct{(}{)}{;}{a}{}{,} 

\usepackage{xcolor}

\begin{document}

   \title{Binary black hole mergers in AGN accretion discs: \\ 
   gravitational wave rate density estimates}

   \subtitle{}
   
       \author{M. Gröbner
          \inst{1,2}
          \and
          W. Ishibashi
          \inst{2}
          \and 
          S. Tiwari
          \inst{2} 
           \and 
          M. Haney
          \inst{2}
          \and  
          P. Jetzer
          \inst{2}
          }

   \institute{Department of Physics, ETH Zürich, Zürich, Switzerland  \label{inst1} \and Physik-Institut, Universität Zürich, Zürich, Switzerland \label{inst2} \\ 
    \email{matthias.groebner@student.ethz.ch, wako.ishibashi@physik.uzh.ch}
 \\
   }

   \date{Received; accepted}

\abstract{
The majority of gravitational wave (GW) events detected so far by LIGO/Virgo originate from binary black hole (BBH) mergers. Among  the different binary evolution paths, the merger of BBHs in accretion
discs of active galactic nuclei (AGNs)  is a possible source of GW detections. We consider an idealised analytical model of the orbital evolution of BBHs embedded in an AGN accretion disc. In this framework, the disc--binary interaction increases the orbital eccentricity and decreases the orbital separation, driving the BBH into a regime where GW emission eventually leads to coalescence. We compute the resulting GW merger rate density from this channel based on a weighted average of the merger timescales of a population of BBHs radially distributed within the AGN accretion disc. The predicted merger rates broadly lie in the range $\mathcal{R} \sim (0.002 - 18) \, \mathrm{Gpc^{-3} yr^{-1}}$. We analyse the dependence of the merger rate density on both the accretion disc and binary orbital parameters, emphasising the important role of the orbital eccentricity. We discuss the astrophysical implications of this particular BBH-in-AGN formation channel in the broader context of binary evolution scenarios. 
}


   \authorrunning{ M. Gröbner et al. }
   \titlerunning{BBH mergers in AGN accretion discs: GW rate density estimates}

   \maketitle



\section{Introduction}

Following the first observation of gravitational waves (GW) in 2015 \citep{Abbott_et_2016_discovery}, GW events are now regularly detected by LIGO/Virgo. The majority of the detections originate from the merger of two stellar-mass black holes or binary black hole (BBH) mergers. Ten confirmed BBH mergers were observed in the first and second observing runs of Advanced LIGO and Advanced Virgo, leading to an estimated BBH merger rate density in the range $\mathcal{R} = (9.7 - 101) \, \, \mathrm{Gpc^{-3} yr^{-1}}$ \citep{Abbott_et_2019_GWTC}, with an updated estimate of $\mathcal{R} = 53.2^{+58.5}_{-28.8} \, \, \mathrm{Gpc^{-3} yr^{-1}}$ (at $90\%$ credibility) \citep{Abbott_et_2019_runs}. 

Two main BBH formation scenarios are currently debated in the literature: isolated evolution in galactic fields  \citep[][and references therein]{Belczynski2020} and dynamical formation in dense stellar environments \citep[][and references therein]{Mandel_Farmer_2018, Mapelli_2018}. 
The predicted merger rates are broadly similar for the two formation channels, but the expected spin and eccentricity distributions are different and may help discriminate between the two paths. For instance, isolated binaries formed in galactic fields are likely to have near-zero eccentricity close to coalescence, while non-negligible eccentricities are expected for binaries formed through dynamical interactions in dense star clusters \citep{Breivik_et_2016, Samsing_Ramirez-Ruiz_2017}. 

In galactic nuclei a large concentration of stellar remnants can be found as a result of dynamical friction and mass segregation.  Indeed, a density cusp of stellar-mass black holes has been recently uncovered in the central parsec of our own Galaxy \citep{Hailey_et_2018}, suggesting that $\sim 10^4$ BHs may be accumulated in the Galactic Centre \citep{Generozov_et_2018}. Galactic nuclei are also known to harbour super-massive black holes (SMBHs) with typical masses of $M_\text{SMBH} \sim (10^6-10^9) M_{\odot}$ \citep[e.g.][and references therein]{Kormendy_Ho_2013}. Accretion of matter from the surrounding environment leads to the formation of an accretion disc around the central SMBH, powering the active galactic nucleus (AGN). Some of the stars orbiting in the nuclear star cluster can then be ground down into the AGN accretion disc, with a significant fraction ending up in binary systems. 

In this context, an alternative formation path for binary black holes was recently discussed in the literature: BBH formation within AGN accretion discs \citep{Stone_et_2017, Bartos_et_2017, McKernan_et_2018}. This novel formation channel has exceptional physical implications: BBH mergers can be considerably accelerated in the gaseous environment of AGN accretion discs and this is the only scenario in which we can plausibly expect electromagnetic counterparts from BBH mergers. The gaseous torques operating in the AGN accretion disc drive the binary into the regime where GW inspiral can take over, eventually leading to coalescence. \cite{McKernan_et_2018} previously assumed that the BBHs merge within the AGN disc lifetime and estimated by this means a merger rate density of $\mathcal{R} = (10^{-3} - 10^4) \, \, \mathrm{Gpc^{-3} yr^{-1}}$. A recent study by \cite{Tagawa2019} combined N-body simulations with semi-analytical tools to investigate the formation, disruption, and evolution of
binaries in AGN discs. Therein, these latter authors found a BBH merger rate density of $\mathcal{R} = (0.02 - 60) \, \, \mathrm{Gpc^{-3} yr^{-1}}$. \\

In this paper we study the evolution of BBHs embedded in AGN accretion discs. A simple analytical model of the disc--binary interaction is established using existing numerical studies as an inspiration. We then couple the disc-driven evolution equations to the corresponding equations of GW-emission. We compute the merger rate density that is expected from this coupled channel based on a weighted average of the merger timescales of a population of BBHs radially distributed in the AGN accretion disc. 
We analyse the physical dependence of the merger rate density on both the binary orbit and the accretion disc parameters, emphasising the important role of the eccentricity. The paper is structured as follows. In Section $\ref{Sect_BBH_in_AGN}$, we summarise the main equations governing the BBH evolution in the disc-driven and GW-driven regimes. In Section $\ref{Sect_BBH_merger_rate}$, we present a detailed computation of the BBH merger rate density estimates for this coupled evolution channel. We analyse the dependence of the resulting merger rate on the underlying physical parameters (Section $\ref{Sect_Results}$). We discuss comparisons with other work and the astrophysical implications in the broader context of BBH formation scenarios in Section $\ref{Sect_Discussion}$.


\section{Binary black
hole evolution in AGN accretion discs}
\label{Sect_BBH_in_AGN}

We briefly introduce an idealised model for the evolution of BBHs embedded in an AGN accretion disc surrounding a central SMBH. We recall the main equations governing the orbital evolution of BBHs throughout the disc-driven and GW-driven regimes. 
We then emphasise the important role of the eccentricity on the orbital evolution in this channel. 

\subsection{The coupled `disc+GW'-driven evolution} \label{modeloutline}

We consider a BBH system on an elliptic orbit characterised by its semi-major axis $a$, its orbital eccentricity $e$, and the total binary mass $M_b = m_1 + m_2$. The total energy of the BBH is given by
$E_b = - \frac{G M_b \mu}{2a}$, 
where $\mu = \frac{m_1 m_2}{M_b} = \frac{q}{(1+q)^2} M_b$ is the reduced mass and $q = m_2/m_1$ is the mass ratio. 
The orbital angular momentum of the binary is given by
$L_b = \mu a^2 \Omega_b \sqrt{1-e^2}$,
where $\Omega_b = \sqrt{\frac{G M_b}{a^3}}$ is the angular frequency of the BBH.

The binary evolves in the gaseous environment of the AGN accretion disc surrounding a central object of mass $M_\text{SMBH}$. We require a stable binary system that is bound to the central mass; that is, when considering the binary system as a whole, orbiting around $M_\text{SMBH}$, the orbital separation $a$ must be smaller than the Roche lobe limit of the central AGN. The stability criterion is given by $\frac{a_{\ast}}{a} \gtrsim \frac{1+e}{1-e_{\ast}} ( \frac{3 M_\textrm{SMBH}}{M_b} )^{1/3}$, where $(a_{\ast}, e_{\ast})$ are the semi-major axis and eccentricity of the outer orbit involving the SMBH \citep{Hoang_et_2018, Fragione_et_2019}. For simplicity, we assume $e_{\ast} = 0$ throughout. The AGN accretion disc is assumed to follow Keplerian rotation, that is, the orbital frequency of the gas at the radius $r$ is $\Omega = \sqrt{\frac{GM_\mathrm{SMBH}}{r^3}}$.  The disc is characterised by its scale-height $H$ and its gas surface density $\Sigma_g = \frac{f_g \sigma^2}{\pi G r}$, where $f_g$ is the gas mass fraction, and where $\sigma$ stands for the stellar velocity dispersion. We relate the central mass $M_\mathrm{SMBH}$ to the stellar dispersion $\sigma$ through the empirical `$M-\sigma$ relation' \citep{Kormendy_Ho_2013}: 
\begin{equation}
\frac{M_\text{SMBH}}{10^9 M_{\odot}} = (0.31^{+0.037}_{-0.033}) \left( \frac{\sigma}{200 \, \mathrm{km s^{-1}} }\right)^{4.38\pm0.29} . 
\end{equation}

The accretion disc is modelled as a geometrically thin and optically thick disc, characterised by the $\alpha$-viscosity prescription $\nu = \alpha c_s H $, where $0 < \alpha < 1$ is a dimensionless parameter and $c_s = H \Omega$ is the local sound speed \citep{Shakura_Sunyaev_1973}. 
We recall that in such a geometrically thin disc the aspect ratio is always $h = H/r \ll 1$. From standard accretion disc theory \citep{Pringle_1981}, the viscous torque acting to transfer angular momentum outwards is given by
\begin{equation}
T_{visc}(r) = - 3 \pi \alpha c_s^2(r) \Sigma_g(r) r^2 . \label{visc}
\end{equation}

Gravitational torques due to the disc--binary interaction may clear an inner cavity in the gas distribution of the AGN disc. This cavity inside of which the BBH resides is surrounded by a circumbinary accretion disc. We assume the tidal torques of the binary to suppress mass accretion into the central cavity, a finding initially noted by \cite{Artymowicz_Lubow_1994} as a result of their smoothed particle hydrodynamic (SPH) simulations. We also assume that neither the primary nor the secondary mass have their own minidiscs. We discuss possible effects of the mass inflow on the orbital evolution in Section \ref{outlook}. Henceforth, we work in the reference frame of the binary system. In this model, no matter is accreting onto the binary and so the binary is injecting angular momentum into the circumbinary disc, and accordingly we require $\dot{L}_b < 0$. We note that according to this convention, in the binary reference frame, the sign of the viscous torque must be opposite to that of Equation \ref{visc}.  By conservation of angular momentum, the binary orbital angular momentum is absorbed by the circumbinary disc and transferred through its inner edge $r_{in}$, thus $ - \dot{L}_b \sim T_{visc} (r_{in})$. 

To relate the viscous angular momentum flux in the inner region of the circumbinary disc to the orbital evolution of the binary, we impose a further assumption to the particular evolution model: we assume that  the complex disc--binary interaction can be solely characterised through the angular momentum of the binary. Concretely, we assume that the binary--disc interaction may well be approximated as an adiabatic process. In addition, we assume that torques act on average axisymmetrically onto the disc. This treatment is also employed for similar evolution models outlined in \cite{Rafikov2013} and \cite{Hayasaki2009}. Based on these two idealised assumptions, the binary energy dissipation can be related to its change in angular momentum through the orbital frequency, that is, $\dot{E}_b = \Omega_b \dot{L}_b$. 

The assumption that non-axisymmetric potential perturbations of the binary system are small around the average binary potential requires that the cavity shape stay circular throughout the evolution. A circular cavity shape may be an acceptable assumption for circular orbits and for orbits with low eccentricities. Non-linear and backcoupling effects are very likely to occur for higher eccentricities and those complicate an analytic treatment considerably. Nevertheless, a circular cavity morphology is adopted in a toy model of \cite{Hayasaki2009}, where the orbital  eccentricity of binaries is allowed to approach $e=1$. Furthermore, gap sizes are commonly reported in terms of the binary orbital semi-major axis up to a multiplicative factor (even for eccentricities up to $0.7$, see $e.g.$ \cite{Artymowicz_Lubow_1994}). On those grounds, we adopt a circular cavity shape throughout. In more realistic situations, the circumbinary disc can be distorted and become eccentric, as a result of its interaction with the central binary \citep{MacFadyen2008,Shi2012}. Eccentric binaries can directly drive the disc eccentricity growth, for example via bar potential \citep{Lubow_Artymowicz_2000}. As observed in three-dimensional magnetohydrodynamic (3D MHD) simulations by \cite{Shi2012}, even near-circular binaries may induce disc eccentricity growth through the impact of gas streams hitting the inner edge of the disc.

Numerical simulations suggest that the inner edge of the circumbinary disc is usually located around twice the semi-major axis, with the cavity size increasing with increasing eccentricity (\cite{Artymowicz_Lubow_1994}, \cite{Hayasaki2007}). In accordance with the above simulations we assume $r_{in} = 2a (1+e)$, and as a consequence the viscous torque is $T_{visc}(r_{in}) = 12 \pi \alpha c_s^2 \Sigma_g a^2 (1+e)^2$.  We further assume that the values of the circumbinary accretion disc parameters can be set equal to the corresponding local values of the background AGN disc \citep[cf.][]{Baruteau_et_2011}. \\

The standard evolution equations of an elliptic Keplerian orbit applied to the thus outlined disc-binary interaction give 
\begin{equation}
\frac{\dot{a}}{a} =  -\frac{24 \pi \alpha c_s^2 \Sigma_g (1+e)^2}{\mu \Omega_b},
\label{Eq_a_disc}
\end{equation}
\begin{equation}
\frac{\dot{e} e}{1-e^2} = \frac{12 \pi \alpha c_s^2 \Sigma_g (1+e)^2 }{\mu \Omega_b} \left(\frac{1}{\sqrt{1-e^2}} -1 \right). 
\label{Eq_e_disc}
\end{equation} 

We note that $\dot{L}_b < 0$ implies $\dot{E}_b < 0$, and therefore in this channel a binary system always evolves toward lower energy. Therefore, even though we have $\dot{e} \geq 0$, the orbit cannot become unbound. 
The trend of binary eccentricity growth ($\dot{e} > 0$) was first found by \citet{Artymowicz_et_1991}, who used SPH simulations to analyse the effects of the circumbinary disc on the evolution of the binary orbital elements. We also note that it would not be meaningful to assert that the disc dynamics alone can provide a full description of the orbital evolution. In fact, a binary evolution premised purely on Equations \ref{Eq_a_disc} and \ref{Eq_e_disc} seems unphysical, for $e \rightarrow 1$ and $a \rightarrow 0$ indicates that the orbit enters a regime where gravitational wave emission cannot be neglected. 

We recall the gravitational-wave-driven evolution equations of the orbital parameters \citep{Peters_1964}:
\begin{equation}
\dot{a} = - \frac{64}{5} \frac{G^3}{c^5} \frac{\mu M_b^2}{a^3} \frac{1}{(1-e^2)^{7/2}} \left( 1 + \frac{73}{24} e^2 + \frac{37}{96} e^4 \right),\label{Eq_a_GW}
\end{equation}
\begin{equation}
\dot{e} = -  \frac{304}{15} \frac{G^3}{c^5} \frac{\mu M_b^2}{a^4} \frac{e}{(1-e^2)^{5/2}} \left( 1 + \frac{121}{304} e^2 \right).
\label{Eq_e_GW}
\end{equation} \\

To investigate the dynamics of the interplay between the 
GW emission and the disc-binary interaction, we combine the disc-driven (Equations \ref{Eq_a_disc}-\ref{Eq_e_disc}) with the corresponding GW-driven (Equations \ref{Eq_a_GW}-\ref{Eq_e_GW}) equations. The coupled `disc+GW' evolution is thus described by
\begin{equation}
\dot{a} = \dot{a}_\text{disc} + \dot{a}_\text{GW} ,
\label{Eq_a_coupled}
\end{equation}
\begin{equation}
\dot{e} =  \dot{e}_\text{disc} + \dot{e}_\text{GW}.
\label{Eq_e_coupled}
\end{equation}  

This coupled system of differential equations can be integrated numerically; this yields the temporal evolution of the binary semi-major axis and orbital eccentricity. Of particular interest is the merger timescale, which is taken as the point where the numerical solution $a(t)$, or equivalently $e(t)$, approaches the abscissa axis. In the following, we consider the coupled `disc+GW'-driven evolution, focusing on the important role of the orbital eccentricity.


\subsection{The role of eccentricity}

\begin{figure}[h!]
  \centering
 \includegraphics[width=0.45\textwidth]{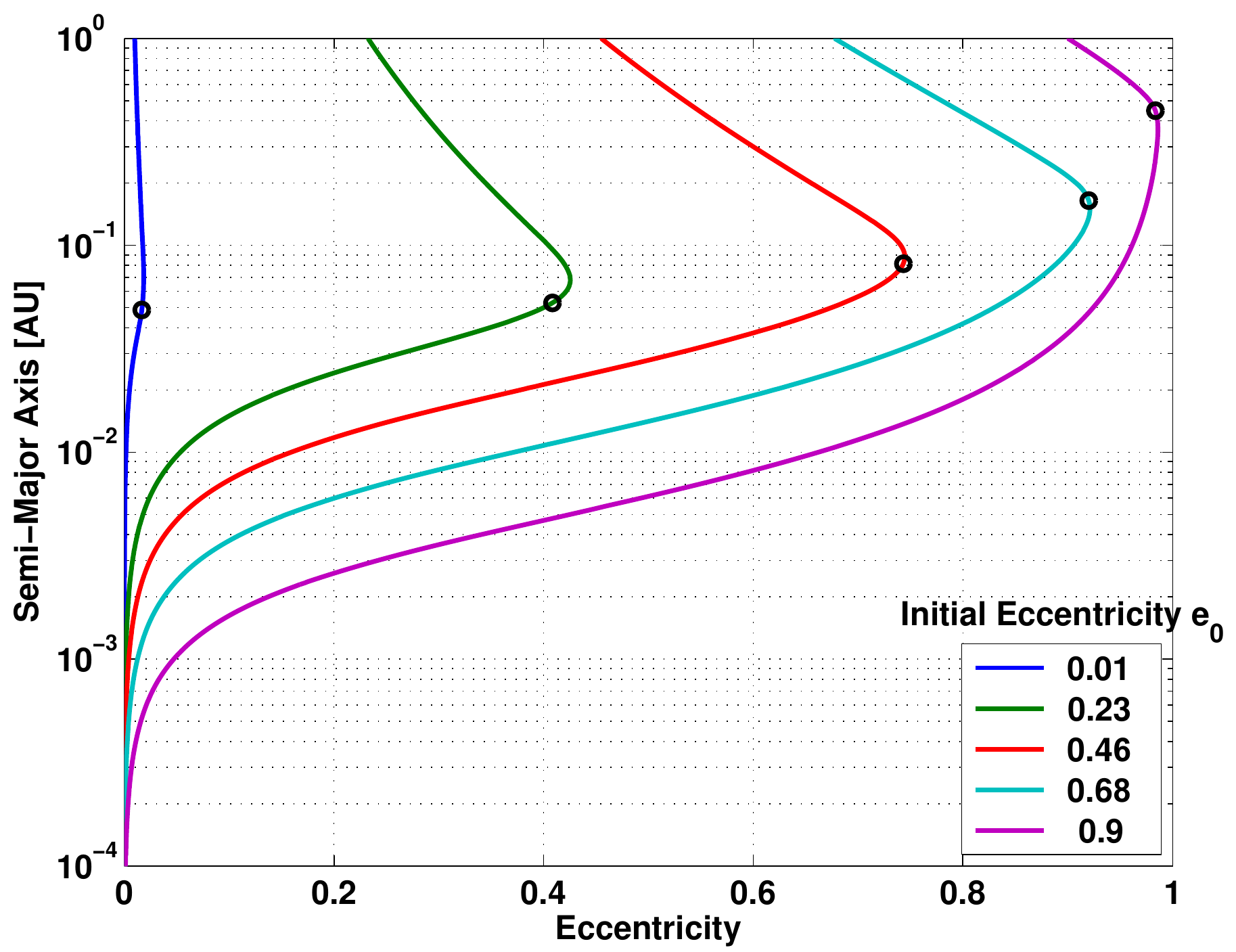}
    \caption{Orbital semi-major axis as a function of the orbital eccentricity for a binary system based on the coupled evolution equations (Equations \ref{Eq_a_coupled}-\ref{Eq_e_coupled}). In this plot we show $a(e)$ for different initial eccentricities ($e_0 = 0.01-0.9$).  }
    \label{evsa_e0}
\end{figure}

Figure \ref{evsa_e0} shows the evolution of the semi-major axis as a function of the orbital eccentricity for different values of the initial eccentricity ($e_0 = 0.01-0.9$).  The black dots indicate the critical radius where GW-emission takes over the disc-driven interaction. We take as fiducial values for the parameters of the accretion disc: viscosity parameter $\alpha = 0.1$, gas fraction $f_g =0.1$, and aspect ratio $h = H/r = 0.01$. The binary parameters are set to: $a_0 = 1$ AU, $M_b = 50 M_\odot$, and $q=1$. In this fiducial model, the binary is located at a distance of $r = 0.1$ pc in an AGN disc surrounding a SMBH of $M_{\text{SMBH}} = 10^7 M_\odot$. We note that the stability condition (cf. Section \ref{modeloutline}) is satisfied for the fiducial model parameters, and so the binary remains bound to the central object. From Figure \ref{evsa_e0} we see that for most values of the semi-major axis, the disc--binary interaction dominates the gravitational-wave emission. Only at the latest stages of inspiral when $a$ is sufficiently low does gravitational emission become important. As for the eccentricity evolution we observe two distinct trends: in the disc-driven regime the orbital eccentricity grows, and in the GW-driven regime the eccentricity decays. The disc--binary interaction shrinks the orbital separation, while at the same time it increases the orbital eccentricity, such as to rapidly bring the BBH into the regime where GW emission takes over. \\

At a critical semi-major axis $a_c$, where $\dot{a}_{\text{disc}}(a_c) = \dot{a}_{\text{GW}}(a_c)$, the binary dynamics evolve from being predominantly disc-driven to predominantly GW-driven; the black circles in Figure \ref{evsa_e0} indicate these transition points. We observe that the critical semi-major axis $a_c$ increases for increasing initial eccentricities. Thus, for smaller $e_0$, the disc interaction drives the binary to lower separations as compared to orbits with higher $e_0$. However, the higher absolute value of the orbital eccentricity at $a_c$ for orbits with higher $e_0$ results eventually in a merging that is faster than for orbits with lower initial eccentricities. We also note that for a small initial eccentricity ($e_0 = 0.01$, blue curve in Figure \ref{evsa_e0}), there is only a small subsequent eccentricity growth in the disc-driven regime. \\

\begin{center}
\begin{figure}[h!]
  \centering
 \includegraphics[width=0.45\textwidth]{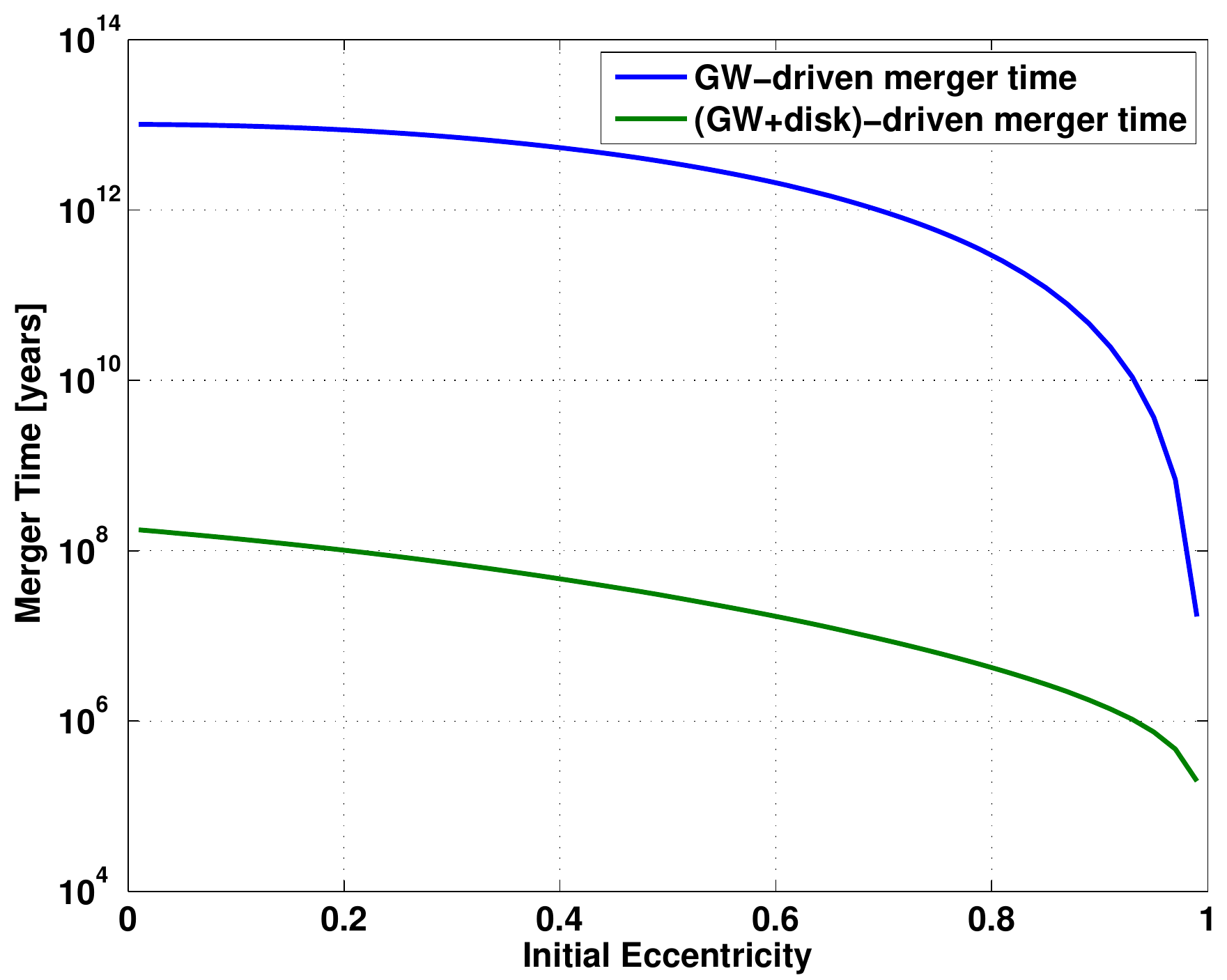}
    \caption{Merger timescales as a function of the initial orbital eccentricity:
    merger times resulting from GW-only decay (blue curve), and both GW+disc evolution (green curve). 
    }
    \label{lifetimes_plot}
\end{figure}
\end{center}

The combined `disc+GW'-driven evolution reduces the purely gravitational wave-driven merger time by several orders of magnitude. Figure $\ref{lifetimes_plot}$ shows, as a function of the initial eccentricity of the  orbit, the purely gravitational wave-driven merger times (blue) and those where disc interactions are also taken into account (green). In the case of purely GW-driven evolution, binaries in this particular configuration would not merge within a Hubble time ($t_H \sim 10^{10}$ yr), unless the initial eccentricity is higher than $e_0 \gtrsim 0.9$. In contrast, in the coupled `disc+GW' evolution, binaries can efficiently merge on timescales of $\tau \sim (10^6 - 10^8)$ yr for all values of the initial eccentricity. For instance, at $e_0 = 0.01$, the merger time is $\tau \sim 10^{13}$ years for the GW-only evolution, while it is $\tau \sim 2 \times 10^8$ years for the combined `disc+GW' evolution.  
Hence, if located in the field, such a low-eccentricity binary would not merge within the age of the universe. A gaseous environment would provide a mechanism for such binaries to merge within a plausible timescale. From Figure $\ref{lifetimes_plot}$ we also observe that in both channels the merger timescale decreases as the initial eccentricity increases: at $e_0 = 0.99$, the merger time is $\tau \sim 2 \times 10^7$ years for the purely GW-driven evolution, compared to $\tau \sim 2 \times 10^5$ years for the coupled disc+GW evolution.

\section{Binary black hole merger rate density estimates} 
\label{Sect_BBH_merger_rate}

In this section, we estimate the BBH merger rate densities that can be expected from the particular evolution channel described in Section \ref{Sect_BBH_in_AGN}. A natural and common way to parametrise the rate density of BBH mergers in AGN discs is \citep[cf.][]{McKernan_et_2018}
\begin{equation}
\mathcal{R} = \frac{n_\text{AGN} N_\text{BH} f_d f_b}{\tau_\text{avg}},\label{rateKernan}
\end{equation}
where $n_\text{AGN}$ is the number density of AGNs (per $\mathrm{Gpc^{-3}}$), $N_\text{BH}$ is the number of stellar-mass BHs in the galactic nucleus (within the central $\sim\text{pc}^3$), $f_d$ is the fraction of BHs embedded in the AGN accretion disc, $f_b$ is the fraction of such BHs residing in binaries, and $\tau_\text{avg}$ is the average BBH merger time. \\

In \cite{McKernan_et_2018} no attempt was made to constrain the actual merger timescale, and binaries were assumed to merge within the AGN disc lifetime ($\tau_\text{AGN} \sim 1-10$ Myr). Based on the model of the previous section, we are able to explicitly compute the merger timescales resulting from the coupled `disc+GW'-driven evolution, taking into account various parameters such as $e_0, a_0, f_g, M_{\text{SMBH}}$, and so on. We modify the prescription of Equation \ref{rateKernan} and estimate the BBH merger rate density from a weighted average of the merger times of a population of BBHs radially distributed within the AGN accretion disc. Below, we provide a description of the computation steps involved in estimating the BBH merger rate density $\mathcal{R}$.

\subsection{Mass distributions}
We combine different mass distributions and mass density profiles to derive an expression for the number of black holes with a given mass range that reside within a radial shell of an AGN disc. 

\subsubsection{Stellar distribution} We start with an initial stellar population. The initial stellar mass distribution seems to follow a multiple power-law function. Here, we adopt the initial stellar mass function of \cite{kroupa02} 
\begin{equation}
f_{\text{IMF}}(m) = \begin{cases}
k_0 \left( \frac{m}{M_\odot} \right)^{-0.3} & \text{if } m < 0.08 M_\odot \\
k_1 \left( \frac{m}{M_\odot} \right)^{-1.3} & \text{if } 0.08  M_\odot  < m < 0.5 M_\odot \\
k_2 \left( \frac{m}{M_\odot} \right)^{-2.3} & \text{if } m > 0.5 M_\odot
\end{cases},
\end{equation}
with the normalisation parameters $k_0,k_1 = 0.08k_0$ and $k_2 = 0.04k_0$. We assume that the lowest mass of the stars $m_\text{s,min}$ is smaller than $0.08 M_\odot$ and that the highest mass $m_\text{s,max}$ is above $0.5 M_\odot$. The total number of stars is $N_* =  \int_{m_\text{s,min}}^{m_\text{s,max}} dm f_{\text{IMF}}(m)$ and the total stellar mass is $M_* = \int_{m_\text{s,min}}^{m_\text{s,max}} dm \, m f_{\text{IMF}}$. For the stellar density profile $\rho_*(r)$, we  adopt the "Nuker law" parametrization (\cite{Lauer1995})
\begin{equation}
\rho_*(r) \propto \left( \frac{r_b}{r} \right)^\gamma \left[ 1 + \left( \frac{r}{r_b}\right)^\alpha \right]^\frac{\gamma - \beta}{\alpha}, 
\end{equation}

and simplify it by using its asymptotic slopes,
\begin{equation}
\rho_*(r) = \begin{cases}
\rho_0 \left( \frac{r}{r_b} \right)^{- \gamma} & \text{if } r \leq r_b \\\rho_0 \left( \frac{r}{r_b} \right)^{- \beta} & \text{if } r > r_b
\end{cases}.
\end{equation}

The break radius $r_b$ is approximately on the order of the radius of influence of the SMBH \citep{Schoedel_et_2018}, characterised as $r_b = GM_{\text{SMBH}} / \sigma^2 $.

\subsubsection{Black hole distribution}
If we assume that every star heavier than $m_{\text{s,cr}} \sim 5 M_\odot$ evolves into a BH, the total number of BHs is 
\begin{align}
N_{\text{BHs}} &= \int_{m_\text{s,cr}}^{m_\text{s,max}} dm \, f_{\text{IMF}} = k M_* , \label{numberBH}
\end{align}
where we define
\begin{equation}
k = \frac{1}{M_\odot} \frac{ - \frac{0.04}{1.3} \left[  \left( \frac{{m_\text{s,max}}}{M_\odot} \right)^{-1.3} -\left( \frac{{m_\text{s,cr}}}{M_\odot} \right)^{-1.3} \right]}{\left[ 0.22 - \frac{1}{1.7} \left( \frac{{m_\text{s,min}}}{M_\odot} \right)^{1.7} - \frac{0.04}{0.3} \left( \frac{{m_\text{s,max}}}{M_\odot} \right)^{-0.3} \right]}.
\end{equation}
The total mass of all BHs is
\begin{align}
M_{\text{BH}} = \int_{m_\text{s,cr}}^{m_\text{s,max}} dm \, m \, f_{\text{IMF}}.
\end{align}

The BH mass distribution is denoted as $f_{\text{BH}}(m)$ and is normalised such that $\int dm f_{\text{BH}}(m) = 1$. We adopt a  power-law mass function for the BH distribution: 
\begin{equation}
f_{\text{BH}}(m) = \xi_0 m^{-\kappa}. \label{IMF}
\end{equation} 
If $N(r)$ denotes the total number of BHs inside some radius $r$, we take the number $dN$ of BHs lying inside $r$ with masses between $m$ and $m + dm$ to be 
\begin{align}
\begin{split}
dN &= N(r)  f_{\text{BH}}(m) dm \\
&= k M_*(r)  f_{\text{BH}}(m) dm, 
\end{split}
\end{align} 
where $M_*(r)$ is the total stellar mass inside radius $r$ ($M_*(\infty) \equiv M_*$). We note that for the last equality we assume that the mass distribution of the BHs is independent of the radius at which they reside. We can write $dN$ in the following form
\begin{equation}
dN = dm \, k  \, f_{\text{BH}}(m)  \int_{0}^r dr \,  \rho_*(r) 4\pi r^2 .
\end{equation}

The number of BHs with masses between $m$ and $m+dm$ in a radial shell from radius $r$ to $r+dr$ is accordingly
\begin{equation}
\begin{cases}
dm \, k \, f_{\text{BH}}(m) \, \frac{4 \pi \rho_0}{3- \gamma} r_b^{\gamma}  \left[ (r+dr)^{3-\gamma} - r^{3-\gamma} \right] & \text{if } r \leq r_b \\
dm \, k \, f_{\text{BH}}(m) \, \frac{4 \pi \rho_0}{3- \beta} r_b^3 \left[   \left(\frac{r+dr}{r_b}\right)^{3-\beta}   -  \left(\frac{r}{r_b}\right)^{3-\beta} \right] & \text{if } r > r_b
\end{cases}. \label{dN/dm/dr}
\end{equation}

\subsection{Rate per galactic nucleus \label{sectionratenucleus}}
To estimate the merger rate per galactic nucleus, we add up the rate contributions coming from different radial shells. We assume that the AGN disc around the SMBH extends from $r_\text{min}$ to $r_\text{max}$. We partition the disc into $\mathcal{I}_r$ intervals. We use $N_{j}$ to denote the number of BHs within a radial shell $[ r_\text{min} + j \Delta r, r_\text{min} + (j+1) \Delta r]$, where $\Delta r = (r_\text{max} - r_\text{min}) /| \mathcal{I}_r|$. The mean radius of a radial shell is $r_j = r_\text{min} + (j+0.5) \Delta r$. The number of BBHs within each radial shell that are embedded in the disc is given by $N_{\text{BBH},j} = N_j f_b f_d / 2$.

\subsubsection{Binary black hole distributions in a radial shell \label{canonicaldistr}}
Within each radial shell labelled `$j$  ' we distribute the number of BBHs $N_{\text{BBH},j}$ for the binary parameter $(m_1,m_2,e_0,a_0) \cong (m_1 \equiv \tilde{M},q,e_0,a_0) $ according to some canonical binary parameter distribution functions. We assume for simplicity that the binary parameter distributions are independent from each other. 
We normalise the distributions in this section with respect to $N_{\text{BBH},j}$.

\paragraph{Primary mass distribution}
We adopt the initial mass function (\ref{IMF}) for the primary mass distribution. 

\paragraph{Secondary mass distribution}
The secondary mass distribution is determined by the distribution of the mass ratio $q$, assumed to follow a uniform distribution
\begin{equation}
f(q) = \vartheta_0, \label{massratiodstrn}
\end{equation}
 where $\vartheta_0$ is a constant and $q_{\text{min}} \leq q \leq q_{\text{max}}$. 
 
\paragraph{Orbital separation distribution}
The distribution of the initial orbital separation $a_0$ is assumed to be logarithmically flat, 
\begin{equation}
f(a_0) = \chi_0 / a_0, \label{orbsepdstrn}
\end{equation}
between the limits $a_{0,\text{min}}$ and $a_{0,\text{max}}$. This distribution is biased towards low semi-major axis.
 
\paragraph{Orbital eccentricity distribution}
We choose the distribution of the initial orbital eccentricity $e_0$ to follow a  thermal distribution, 
\begin{equation}
 f(e_0) = \varsigma_0 e_0, \label{eccentdstrn}
\end{equation}
between the limits $e_{0,\text{min}}$ and $e_{0,\text{max}}$. The number of BBHs below a given eccentricity $e$ scale in a thermal distribution as $e^2$; a thermal distribution thus favours high eccentricities.

\subsubsection{Calculation of rate within a radial shell} 
In order to calculate the rate of coalescences in a radial shell $j$, we partition each binary parameter into bins and distribute $N_{\text{BBH},j}$ among the different bins. The merger rate is then determined as follows: \\*

The quantities $\mathcal{I}_{x}$ (where $x$ is one of $\tilde{M},q,e_0,$ or $a_0$) will denote the number of bins for the corresponding binary distributions $x$. We use  $N_{x,i}$ to denote the number of BBHs residing within the bin $i \in \mathcal{I}_x$ of the binary distribution $x,$ and use  $x_i$ to denote the average of the variable $x$ within bin $i$. We need to calculate the merger time $\tau_{\text{merger},\lambda}$ for
all possible combinations of $\lambda \in \mathcal{I}_{\tilde{M}} \times
\mathcal{I}_q \times \mathcal{I}_{a_0} \times \mathcal{I}_{e_0}$  . We take  the average quantities $x_i$ of the orbital parameter $x$ in bin $i$ as the input values needed for the numerical calculation of the merger time. The total merger rate resulting from a radial shell $j$ is therefore
\begin{equation}
\mathcal{R}_j = \sum_{\lambda} \frac{N_{\lambda}}{N_{\text{BBH},j}^3} \frac{1}{\tau_{\text{merger},\lambda}}, \label{mergerrateshell}
\end{equation}
where the sum goes over all $\lambda \in \mathcal{I}_{\tilde{M}} \times \mathcal{I}_q \times \mathcal{I}_{a_0} \times \mathcal{I}_{e_0}$ and where $N_{\lambda}$ is the number of BBHs residing with the binning configuration $\lambda$. 
 
 \subsubsection{Merger rate for a galactic nucleus}
The merger rate for a galactic nucleus is obtained by summing up the rate ($\ref{mergerrateshell}$) for each radial shell $j$ 
\begin{equation}
\mathcal{R}_{\text{gal}} = \sum_{j \in \mathcal{I}_r} \mathcal{R}_j.
\end{equation}

We operate with the binning $\mathcal{I}_{r} = \mathcal{I}_{e_0} = 20$ and $\mathcal{I}_{\tilde{M}} = \mathcal{I}_{q} = \mathcal{I}_{a_0} = 10$, as it is both stable with respect to finer binnings and computationally economical.

 \subsection{Merger rate density \label{sectionratedensity}}

The merger rate density $\mathcal{R}$ is given by
\begin{equation}
\mathcal{R} =  \mathcal{R}_{\text{gal}} \, n_{\text{AGN}}. \label{rate2}
\end{equation}

We note that the number $N_{*}$ of stars within $r_{\text{max}}$, the fraction $f_b$ of BH in BBHs, the fraction $f_d$ of BHs embedded in the disc, and the number density $n_{\text{AGN}}$ of AGNs affect the merger rate density $\mathcal{R}$ only as scaling factors, and we report rate densities in a form where we set these to a fixed value. \\

The number density of AGNs is given by $n_\text{AGN} = f_\text{AGN} N_\text{GN}$, where $f_\text{AGN}$ is the fraction of galactic nuclei that are active, and $N_\text{GN}$ is the average number density of galactic nuclei. The lower limit on $n_\text{AGN}$ would correspond to luminous quasars ($f_\text{AGN} \sim 0.01$), while the upper limit corresponds to the more common low-luminosity AGNs ($f_\text{AGN} \sim 0.3$). A significant fraction of BHs in galactic nuclei reside in binary systems, with a typical binary fraction in the range $f_b \sim (0.01-0.2)$ \citep{McKernan_et_2018}, and possibly up to $f_b \sim (0.6-0.8)$ for binaries located in the innermost regions of AGN gaseous discs \citep{Secunda_et_2019}. 
The fraction of BBHs that end up embedded in the AGN disc scales with the disc aspect ratio and is roughly given by $f_d \geq H/r$, covering the range $f_d \sim (0.01-0.7)$ for geometrically thin/thick discs \citep{Ford_McKernan_2019}. 
We therefore adopt the following values for the population parameters (indicating in parenthesis the lower bound and upper bound): 
\begin{itemize}
\item[$\diamond$] $f_b = 0.1$ [$0.01,0.2$],
\item[$\diamond$] $f_d = 0.01$ [$0.01,0.7$],
\item[$\diamond$] $n_{\text{AGN}} = 4 \times 10^4$ Gpc$^{-3}$ [$4 \times 10^4$ Gpc$^{-3}$, $3 \times 10^{6}$ Gpc$^{-3}$].
\end{itemize} 

In addition we fix the maximal radius as $r_{\text{max}} = 1$ pc. The inner edge of the accretion disc can in principle extend down to the innermost stable circular orbit, but we do not expect binaries to reside there. 
We therefore set fiducially $r_{\text{min}}$ to 100 times the Schwarzschild radius of the SMBH. We estimate the number $N_{*}$ of stars residing within \mbox{$1$ pc}  as $1 \times 10^6$. Furthermore, we use the fiducial values $m_{\text{s,min}} = 0.01 M_{\odot}, m_{\text{s,max}} = 200 M_{\odot}$ , and $m_{\text{s,cr}} = 5 M_{\odot}$. The number of BHs within \mbox{$1$ pc}, obtained from formula (\ref{numberBH}) is then about $7.5 \times 10^3$, which is close to what is observed in our Galactic Centre \citep{Generozov_et_2018} and to the lower end of the range $N_\text{BH} \sim (10^4-10^6)$ estimated in \citet{McKernan_et_2018}. \\

Additionally,  throughout we use the fiducial values $\beta = 3.2$ and $\gamma = 1.5$ for the mass distribution parameters. Also, unless subject to variation, we adopt a viscosity parameter $\alpha = 0.1$, a gas fraction $f_g = 0.1$, and a disc aspect ratio \mbox{$H/r = 0.01$}.   
We assume $M_{\text{SMBH}} = 1 \times 10^7 \, M_\odot $ for our fiducial model, and adopt for the BH mass distribution a Salpeter mass function, that is, we set $\kappa = 2.35$. Unless a binary parameter is subject to variation, we use the respective distributions of Section (\ref{canonicaldistr}) and the following ranges of the orbital parameters 
\begin{itemize}
\item[$\diamond$]  $[m_{\text{BH,min}},m_{\text{BH,max}}] = [5, 50] \, M_{\odot}$,
\item[$\diamond$]  $[q_\text{min},q_{\text{max}}] = [0.1,1]$,
\item[$\diamond$]  $[a_{0,\text{min}},a_{0,\text{max}}] = [1, 100]$ AU,\item[$\diamond$] $[e_{0,\text{min}},e_{0,\text{max}}] = [0.001, 0.99] $ . \\
\end{itemize}


\section{Results: GW event rates}
\label{Sect_Results}

\begin{table}
\begin{tabular}{| p{2.55cm} | p{1.65cm} | p{1.65cm} | p{1.5cm} |}
\hline
initial eccentricity & thermal \newline distribution between $[0.001, 0.99]$ & uniform distribution between $[0.001, 0.99]$ &  $\delta$-function $e_0$=0.001  \\  \hline  \hline
$\mathcal{\tau_{\text{avg}}} \,\, [\text{years}]$ & $2.4 \times 10^{8}$ & $ 2.5 \times 10^8$ & $1.8 \times 10^9$   \\  \hline
$\mathcal{\tau_{\text{med}}} \,\, [\text{years}]$ & $7.0 \times 10^{6}$ &  $  7.0 \times 10^6$ & $1.5 \times 10^9$   \\  \hline
$\mathcal{R} \, \, [\text{year}^{-1} \, \text{Gpc}^{-3}]$ & $0.18$ & $9.5 \times 10^{-2}$ & $ 3.0 \times 10^{-6}$   \\  \hline
\end{tabular}

\bigskip

\begin{tabular}{| p{2.5cm} | p{2.2cm} | p{2.9cm} |}
\hline
orbital evolution  & GW driven & GW and disc driven \\  \hline \hline
$\mathcal{\tau_{\text{avg}}} \,\, [\text{years}]$ & $3.9 \times 10^{21} $ & $ 2.4 \times 10^8$    \\  \hline
$\mathcal{\tau_{\text{med}}} \,\, [\text{years}]$ & $2.5 \times 10^{19}$ &  $7.0 \times 10^6$    \\  \hline
$\mathcal{R} \,\, [\text{year}^{-1} \, \text{Gpc}^{-3}]$ & $5.5 \times 10^{-10}$ & $0.18$    \\  \hline
\end{tabular} 

\bigskip

\begin{tabular}{| p{3.2cm} | p{2.2cm} | p{2.2cm} |}
\hline
initial semi-major axis & log-flat \newline distribution between $[0.1, 100]$ AU & uniform \newline distribution between $[0.1, 100]$ AU   \\  \hline \hline
$\mathcal{\tau_{\text{avg}}} \,\, [\text{years}]$ & $2.4 \times 10^8$ & $ 2.4 \times 10^8$   \\  \hline
$\mathcal{\tau_{\text{med}}} \,\, [\text{years}]$ & $7.0 \times 10^6$ &  $6.6 \times 10^6$    \\  \hline
$\mathcal{R} \,\, [\text{year}^{-1} \, \text{Gpc}^{-3}]$ & $0.18$ & $ 0.49$   \\  \hline
\end{tabular} 

\bigskip

\begin{tabular}{| p{2.4cm} | p{1.6cm} | p{1.6cm} | p{1.6cm} |}
\hline
$\kappa$  & 1.35 & 2.35 & 2.85 \\  \hline \hline
$\mathcal{\tau_{\text{avg}}} \,\, [\text{years}]$ & $ 2.4 \times 10^8 $ & $ 2.4 \times 10^8$ & $2.4 \times 10^8$   \\  \hline
$\mathcal{\tau_{\text{med}}} \,\, [\text{years}]$ & $7.0 \times 10^6$ &  $7.0 \times 10^6$ & $6.9 \times 10^6$   \\  \hline
$\mathcal{R} \,\, [\text{year}^{-1} \, \text{Gpc}^{-3}]$ & $0.12$ & $0.18$ & $0.20$   \\  \hline
\end{tabular}

\bigskip

\begin{tabular}{| p{2.4cm} | p{1.6cm} | p{1.6cm} | p{1.6cm} |}
\hline
 $M_{\text{SMBH}}$ & \centering $1 \times 10^6 \, M_\odot$ & \centering $1 \times 10^7 \, M_\odot$  & $1 \times 10^8 \, M_\odot$    \\  \hline  \hline
$\mathcal{\tau_{\text{avg}}} \,\, [\text{years}]$ & $ 3.2 \times 10^9$ & $ 2.4 \times 10^8$ & $1.8 \times 10^{7}$   \\  \hline
$\mathcal{\tau_{\text{med}}} \,\, [\text{years}]$ & $ 1.7 \times 10^8$ &  $7.0 \times 10^6$ & $ 2.7 \times 10^{5}$   \\  \hline
$\mathcal{R} \,\, [\text{year}^{-1} \, \text{Gpc}^{-3}]$ & $4.0 \times 10^{-2}$ & $0.18$ & $0.75$   \\  \hline
\end{tabular}

\bigskip

\begin{tabular}{| p{2.4cm} | p{1.6cm} | p{1.6cm} | p{1.6cm} |}
\hline
$\alpha$  & 0.01 & 0.1 & 0.3 \\  \hline \hline
$\mathcal{\tau_{\text{avg}}} \,\, [\text{years}]$ & $ 1.4 \times 10^9 $ & $ 2.4 \times 10^8$ & $1.0 \times 10^8$   \\  \hline
$\mathcal{\tau_{\text{med}}} \,\, [\text{years}]$ & $6.2 \times 10^7$ &  $7.0 \times 10^6$ & $2.4 \times 10^6$   \\  \hline
$\mathcal{R} \,\, [\text{year}^{-1} \, \text{Gpc}^{-3}]$ & $1.8 \times 10^{-2} $ & $0.18$ & $0.53$   \\  \hline
\end{tabular} 

\bigskip

\begin{tabular}{| p{2.4cm} | p{1.6cm} | p{1.6cm} | p{1.6cm} |}
\hline
$f_g$  & 0.05 & 0.1 & 0.5 \\  \hline \hline
$\mathcal{\tau_{\text{avg}}} \,\, [\text{years}]$ & $ 4.1 \times 10^8 $ & $ 2.4 \times 10^8$ & $ 6.9 \times 10^7$   \\  \hline
$\mathcal{\tau_{\text{med}}} \,\, [\text{years}]$ & $1.4 \times 10^7$ &  $7.0 \times 10^6$ & $ 1.5 \times 10^6$   \\  \hline
$\mathcal{R} \,\, [\text{year}^{-1} \, \text{Gpc}^{-3}]$ & $8.8 \times 10^{-2} $ & $0.18$ & $0.88$   \\ \hline 
\end{tabular}

\bigskip

\begin{tabular}{| p{2.4cm} | p{1.6cm} | p{1.6cm} | p{1.6cm} |}
\hline
$ h = H/r$  & 0.001 & 0.01 & 0.1 \\  \hline \hline
$\mathcal{\tau_{\text{avg}}} \,\, [\text{years}]$ & $ 8.5 \times 10^7 $ & $ 2.4 \times 10^8$ & $ 6.8 \times 10^6$   \\  \hline
$\mathcal{\tau_{\text{med}}} \,\, [\text{years}]$ & $5.4 \times 10^8$ &  $7.0 \times 10^6$ & $ 7.9 \times 10^4$   \\  \hline
$\mathcal{R} \,\, [\text{year}^{-1} \, \text{Gpc}^{-3}]$ & $1.8 \times 10^{-3} $ & $0.18$ & $17.6$   \\  \hline
\end{tabular} \\* \\*
\caption{Merger rate densities and merger timescales for variations in the binary orbit parameters and AGN disc parameters.}
\label{tablerates}
\end{table}

In Table \ref{tablerates}, we display the resulting BBH merger rate densities $\mathcal{R}$ alongside the average and median merger timescales ($\tau_\text{avg}, \tau_\text{med}$) for variations in the binary orbit and AGN disc parameters. 
We note that the average and median merger timescales generally differ by about two orders of magnitude. This attests to the inadequacy of resorting to average merger timescales for rate estimates. Resorting to an average timescale when determining rates leads to underestimation of BBH mergers with merger times smaller than $\tau_\text{avg}$ and overestimation of BBHs with merger times higher than $\tau_\text{avg}$. \\

Overall, our predicted BBH-in-AGN merger rates broadly lie in the range $\mathcal{R} \sim (0.002 -18)  \, \mathrm{Gpc^{-3} yr^{-1}}$, with a characteristic value of $\mathcal{R} \sim 0.2 \, \mathrm{Gpc^{-3} yr^{-1}}$ for the fiducial parameter choice described in the previous section. This is to be compared with the corresponding rate of $\mathcal{R} \sim 5.5 \times 10^{-10} \, \mathrm{Gpc^{-3} yr^{-1}}$ expected for BBH mergers driven solely by GW emission.  Thus the merger rate seems to be astrophysically irrelevant in the case of purely GW-driven inspiral, while taking into account a disc--binary interaction can yield astrophysically significant rates. This again highlights the importance of including the disc-driven regime in the coupled `disc+GW' evolution scenario.

In addition, the eccentricity distribution plays an important role in determining the BBH merger rates (see Table \ref{tablerates}). 
For instance, assuming that all binaries are initially on quasi-circular orbits ($\delta$-function at $e_0 = 0.001$), the resulting merger rate density is $\mathcal{R} \sim 3 \times 10^{-6} \, \mathrm{Gpc^{-3} yr^{-1}}$. This astrophysically low value may also be related to the fact that for near-zero initial eccentricities, there is negligible eccentricity growth in the disc-driven regime (cf. Figure \ref{evsa_e0}). Considering a more realistic distribution of initial eccentricities, such as the thermal or uniform distributions, leads to BBH merger rates that are $\gtrsim 4$ orders of magnitude higher. We assume a thermal distribution for our fiducial model. Assuming a uniform distribution instead results in a merger rate that is a factor of  approximately two lower than the fiducial choice. This can be attributed to the fact that a thermal distribution tends to favour high-eccentricity binaries (the mean eccentricity in a thermal distribution is two-thirds, while it is one-third in a uniform distribution). 

Similarly, a logarithmically flat distribution in the initial orbital separation is biased toward lower semi-major axis compared to a uniform distribution.  As the orbital decay is more efficient at large separations in the disc-driven regime, the resulting merger rate is higher in the latter case (cf. Table \ref{tablerates}). On the other hand, variations in the power-law slope of the BH mass distribution seem to have only a modest effect on the merger rates (within a plausible range of $\kappa$). 

With regards to the AGN parameters, we see that the merger rate is an increasing function of the SMBH mass: $\mathcal{R}$ is higher for binaries orbiting more massive central objects. Likewise, we observe that increasing the viscosity parameter $\alpha$, gas fraction $f_g$, and disc aspect ratio $h$ lead to higher BBH merger rates. Among the accretion disc parameters, the aspect ratio seems to be the most important factor determining the global $\mathcal{R}$ range. In physical terms, the predicted trends can be interpreted as follows: in the disc-driven regime, both the orbital decay and eccentricity growth scale as $\propto \alpha f_g h^2$. Therefore, binaries embedded in a gas-rich viscous accretion disc with a large aspect ratio are more likely to merge rapidly, which is reflected in higher GW event rates.


\section{Discussion}
\label{Sect_Discussion}

\subsection{The coupled `disc+GW' evolution of BBHs in AGN discs }

We consider the evolution of BBHs embedded in AGN accretion discs governed by the coupled `disc+GW'-driven evolution equations (Equations \ref{Eq_a_coupled}-\ref{Eq_e_coupled}). It is well known that in the GW-driven regime the binary orbital decay is more efficient at small semi-major axis. According to our model, the orbital decay in the disc-driven regime is instead more efficient at large separations. There are two interesting physical trends by which the disc-binary interaction facilitates the merging of the BBH: orbital decay and eccentricity growth. On the one hand, gaseous torques act to shrink the binary separation to small enough radii where GW inspiral can take over. On the other hand, there can be considerable eccentricity growth in the disc-driven regime. Consequently the binary enters the GW-driven regime with high eccentricity, where the merging is fostered due to the steep dependence of the orbital decay on the eccentricity (Equation \ref{Eq_a_GW}). Both effects combine to accelerate the BBH merger in the coupled `disc+GW' evolution scenario, as compared to a purely GW-driven evolution. 

As a result, binaries that otherwise would not merge within the age of the universe (if located in galactic fields), can be efficiently driven to undergo a merger in the gaseous environment of AGN accretion discs. This observation is not new and has been suggested by various authors \citep{Cuadra_et_2009,McKernan_et_2018}. A novelty of this work lies in the quantification of this statement by explicitly computing the resulting BBH merger rates based on the weighted averages of binary populations embedded within the AGN accretion disc. 
From our calculations we expect typical GW event rates in the range $\mathcal{R} \sim (0.002 -18)  \, \mathrm{Gpc^{-3} yr^{-1}}$. The current LIGO-Virgo estimate for the BBH merger rate density lies in the range $\mathcal{R} = (9.7 - 101) \, \mathrm{Gpc^{-3} yr^{-1}}$. 
Our predicted merger rates are clearly lower than the observationally inferred ones. Importantly, not all black hole binaries originate from this particular AGN formation channel. Nevertheless, this peculiar path may significantly contribute to the total merger rate. In fact, our rate estimates should be viewed as lower limits, since we adopt the lower bounds for most of the physical parameters (e.g. $n_\text{AGN}, f_d$, see Section \ref{Sect_BBH_merger_rate}). Therefore the actual BBH merger rates resulting from gas-assisted orbital evolution could be easily one or more orders of magnitudes higher. \\

For comparison, a merger rate of $\mathcal{R} \sim 3 \, \mathrm{Gpc^{-3} yr^{-1}}$ is expected for BBHs embedded in a self-gravitating disc of AGNs \citep{Stone_et_2017}; while a similar rate of $\mathcal{R} \sim 1.2 \, \mathrm{Gpc^{-3} yr^{-1}}$ is predicted for BBHs trapped in the innermost regions of the AGN disc \citep{Bartos_et_2017}. Nevertheless, large uncertainties are always involved, and taking into account the entirety of the allowed range of physical parameters leads to a span of $\mathcal{R} = (10^{-3} - 10^4) \, \mathrm{Gpc^{-3} yr^{-1}}$ for BBH mergers in AGN discs \citep{McKernan_et_2018}. 
 Recently, \cite{Tagawa2019} combined N-body simulations with a semi-analytical model to investigate how binaries form and merge in AGN discs. The setup of these latter authors incorporates the formation and disruption of binaries, gas dynamical friction, torques from circumbinary discs, migration in the AGN disc, and several different types of stellar interaction. By this means they estimate a merger fraction per BH in an AGN lifetime, and thereby find a volumetric BBH merger rate of $\mathcal{R} = (0.02 - 60) \, \mathrm{Gpc^{-3} yr^{-1}}$. In contrast to this latter study, we do not assume zero orbital eccentricity in our evolution channel, although indeed our treatment rests on idealised assumptions (see Sections \ref{modeloutline} and \ref{outlook}). From our toy model, we explicitly compute the weighted average of the merger timescales of BBH populations radially distributed in the AGN accretion disc, and calculate the resulting GW event rates. Overall, our predicted merger rates are broadly comparable with similar scenarios of BBHs in AGN discs, although our fiducial rate lies on the lower side (but we also note that a favourable combination of the accretion disc parameters could easily increase the fiducial rate, bringing it more in line with other estimates). 

One of the purposes of this study is to explore how the rate density $\mathcal{R}$ varies as a function of the underlying physical parameters, that is, the binary and accretion disc configurations. For what concerns the accretion disc parameters, we observe that higher viscosities, larger gas fractions, and larger aspect ratios can lead to enhanced merger rates. According to our parameter space study, the key factor is the disc aspect ratio $h = H/r$, with geometrically thicker discs ($h \sim 0.1$, that is not razor-thin) allowing larger BBH merger rates. As already noted, the merger rate is also a slowly increasing function of the central super-massive black hole mass $M_\text{SMBH}$. 
Summarising, in our simplified model, BBH mergers are most effective when the binaries are embedded in a high-viscosity, gas-rich, and relatively thick accretion disc surrounding a massive central object. \\

A distinctive feature of the BBH evolution in AGN accretion discs in this channel  is the eccentricity growth in the disc-driven regime. The orbital eccentricity plays a major role in this scenario: it is responsible for considerably reducing the merger timescales, thereby increasing the corresponding merger rate densities. Importantly, including eccentric binaries leads to GW event rates enhanced by several orders of magnitude with respect to circular binaries (Section \ref{Sect_Results}). We note that a finite non-zero value of the initial eccentricity is also required in order to get efficient eccentricity growth in the disc regime (Figure \ref{evsa_e0}). Such initial seed eccentricities could be acquired through different stellar dynamical processes operating in galactic nuclei, such as the Kozai-Lidov mechanism, and other triple systems formed via binary--binary or binary--single interactions \citep[e.g.][and references therein]{Rasskazov_Kocsis_2019}. \\

Another unique signature of a BBH-in-AGN channel is the potential association with electromagnetic (EM) counterparts \citep{Stone_et_2017, Bartos_et_2017}. In general, BBH mergers are not expected to produce EM counterparts, unlike mergers involving neutron stars. However, in the dense gaseous environment of AGN accretion discs, binary black holes may accrete significant amounts of gas at high rates, possibly exceeding the Eddington limit. Such super-Eddington accretion episodes may lead to transient EM counterparts, through the development of ultrafast outflows and/or relativistic jets, radiating over a broad range of wavelengths \citep[e.g.][]{Murase_et_2016}. The associated high-energy emission, such as X-rays or gamma-rays, could potentially be detected by current or upcoming space-borne instruments \citep{Bartos_et_2017}. \\

\subsection{ Caveats and outlook} \label{outlook}

Finally, because many simplifying assumptions are adopted in this work, our results should be considered as tentative. As already mentioned in Section \ref{modeloutline}, our disc--binary interaction is mainly based on angular momentum conservation. A more refined and correct treatment of the problem requires taking into account resonant interaction mechanisms, such as co-rotation and Lindblad resonances, between the binary and its surrounding disc. The outer Lindblad resonances usually tend to increase the binary eccentricity, whereas the inner Lindblad and corotational resonances tend to damp the eccentricity growth \citep{Artymowicz_et_1991}. At high eccentricities, several competing resonances can be simultaneously present. Around $e \gtrsim 0.5$, the opposite effects of such resonances may partially cancel out, considerably reducing the eccentricity growth rate \citep{Lubow_Artymowicz_2000}. 
Moreover, the eccentricity pumping tends to decline, and eventually saturates for higher eccentricities  \citep{Dermine2013}.  
Due to the potential overestimation of the eccentricity growth rate, our reported merger timescales and GW rate densities may be too high overall. Further work is required to see how such non-linear effects can be properly incorporated into our toy model (one possibility is to use the resonance formalism of \cite{goldreich}, but this formalism is inapplicable unless $e \ll 1$).

Furthermore, we assume that the binary torque is efficient at suppressing the mass inflow from the disc. That is, we completely neglect gas inflows into the cavity within which the binary resides, and the subsequent accretion onto the individual black holes. 
 Based on this, in our model the binary is only losing angular momentum to the disc and is thus invariably migrating inwards. More generally, the binary loses angular momentum to the disc via gravitational torques, while it gains angular momentum through accretion. The binary orbital decay rate is then roughly determined by the balance between these two opposite processes. 
Several numerical studies suggest that the trend of inward migration holds even when including the effects of mass transfer, because the angular momentum gain by accretion is usually not enough to offset the angular momentum loss to the disc \citep{MacFadyen2008,Cuadra_et_2009,Shi2012}. Three-dimensional MHD simulations of circumbinary discs indicate the development of two narrow streams flowing from the inner edge of the disc towards the binary, supplying substantial mass accretion. The time-averaged accretion rate onto the binary is found to be comparable to that on a single central mass \citep{Shi_Krolik_2015}. The large accretion rates imply a smaller orbital shrinkage rate, although the net result is $\dot{a}/a < 0$ \citep{Shi2012}. 

In contrast to the above conclusions, recent 2D and 3D viscous hydrodynamical simulations of circumbinary accretion indicate that accreting binaries consistently gain angular momentum from the disc \citep{Moody2019, Munoz_et_2019, Munoz_et_2020}. This implies that the binaries expand as they accrete, with positive $\langle \dot{a} \rangle > 0$ (for moderate mass ratios). 
On the other hand, \citet{Ragusa_et_2016} argue that in the case of thin discs with small aspect ratios ($h \lesssim 10^{-2}$, typical of AGN discs) the accretion rate is suppressed in their 3D SPH simulations (note that most numerical studies rather assume $h \sim 0.1$). 
A similar conclusion, namely that the accretion rate onto the binary is significantly reduced for thin discs, is also reached in 2D hydrodynamic simulations  \citep{Terquem_Papaloizou_2017}. Such a dependence on the disc aspect ratio was previously noted by \citet{Artymowicz_Lubow_1996}, who suggested that the development of efficient accreting gas streams requires warm and thick discs (with aspect ratio $h \gtrsim 0.05$). 
The overall binary evolution is also influenced by the uncertain evolution of the eccentricity, which may assist the orbital decay \citep{Duffell_et_2019}, but further investigations are required to obtain a more complete picture.

In addition, we do not consider the stellar processes occurring on larger scales. These are also likely at the origin of the non-negligible initial eccentricities, which are relevant for the subsequent binary evolution. The inclusion of such additional mechanisms should lead to more physically motivated initial conditions.  Ideally, one would follow the evolution of the BBHs all the way from galactic scales down into the GW-driven regime.
Indeed, BBHs are ultimately driven to merger by a combination of at least three physical processes operating on different physical scales: three-body stellar scatterings at large radii, gaseous torques in AGN accretion discs at intermediate radii, and finally GW emission at small radii. \\

\begin{acknowledgements}
WI and PJ acknowledge support from the University of Zurich.  ST is supported by Forschungskredit Nr. FK-19-114.  MH acknowledges support from Swiss National Science Foundation (SNSF) grant Nr. IZCOZ0-177057.
\end{acknowledgements}

\bibliographystyle{aa}
\bibliography{biblio_bis}

\end{document}